\documentclass[a4paper,fleqn,usenatbib]{mnras}

\usepackage{mathptmx}

\usepackage[T1]{fontenc}
\usepackage{ae,aecompl}

\usepackage{graphicx}	
\usepackage{amsmath}	
\usepackage{amssymb}	
\usepackage[usenames,dvipsnames]{xcolor}

\newcommand{\lta}{\mathrel{\hbox{\raise 0.6 ex \hbox{$<$}\kern
                   -1.8 ex\lower .5 ex\hbox{$\sim$}}}}                
\newcommand{\gta}{\mathrel{\hbox{\raise 0.6 ex \hbox{$>$}\kern
                   -1.7 ex\lower .5 ex\hbox{$\sim$}}}}
                   
\DeclareMathAlphabet{\mathsc}{T1}{cmr}{m}{sc}

\title[Element distributions, assuming a mass loss]{Time-dependent atomic diffusion in
  the atmospheres of CP stars. A big step forward: introducing numerical models including 
  a stellar mass loss}

\author[G. Alecian \& M.J. Stift]
{
G.~Alecian$^{1}$\thanks{E-mail:georges.alecian@obspm.fr},
M.J.~Stift$^{2,3}$
\\
$^{1}$LUTH, Observatoire de Paris, PSL Research University, CNRS, Universit{\'e} Paris
Diderot, 5 place Jules Janssen, F-92190 Meudon, France\\
$^{2}$Armagh Observatory, College Hill, Armagh BT61 9DG, UK\\
$^{3}$Kuffner-Sternwarte, Johann Staud-Strasse 10, A-1160 Wien, Austria\\
}

\date{Accepted 2018 November 1. Received 2018 October 16; in original form 2018 August 30}

\pubyear{2018}

\begin{document}
\label{firstpage}
\pagerange{\pageref{firstpage}--\pageref{lastpage}}
\maketitle

\begin{abstract}
  Calculating abundance stratifications in ApBp/HgMn star atmospheres, we are
  considering mass-loss in addition to atomic diffusion in our
  numerical code in order to achieve more realistic models. These numerical
  simulations with mass-loss solve the time dependent continuity equation for
  plane-parallel atmospheres; the procedure is iterated until stationary
  concentrations of the diffusing elements are obtained throughout a large part
  of the stellar atmosphere. We find that Mg stratifications in
  HgMn star atmospheres are particularly sensitive to the presence of a
  mass-loss. For main-sequence stars with $T_{\rm{eff}}\approx
  12000$\,K, the observed systematic mild underabundances of this element can be
  explained only if a mass-loss rate of around $4.2\,10^{-14}$
  solar mass per year is assumed in our models. Numerical simulations also
  reveal that the abundance stratification of P observed in the HgMn star
  HD\,53929 may be understood if a weak horizontal magnetic field of about 75\,G
  is present in this star. However, for a better comparison of our results with
  observations, it will be necessary to carry out 3D modelling, especially when
  magnetic fields and stellar winds -- which render  the atmosphere anisotropic
  -- are considered together.

\end{abstract}




\section{Introduction}
\label{intro}

The modelling of chemically peculiar (CP) star atmospheres including atomic
diffusion constitutes a long-standing challenge. Between the first paper by
\citet{MichaudMi1970y} and the most recent calculations of
\citet{AlecianAlSt2017}, great progress has been made. However, as pointed out
in the conclusions of \citet{MichaudMiAlRi2015} and in several theoretical works
on this subject, detailed observed abundances of individual CP stars are still
difficult to reproduce in numerical models, even in the case HgMn stars which
are considered as the simplest ones to model. Theses difficulties are partly due
the fact that atomic diffusion is a very slow process, therefore very sensitive
to any perturbation due to other physical processes such as macroscopic motions
(even weak turbulence, convection, wind or mass-loss), magnetic
fields, etc. when models of inhomogeneous elements distributions are computed in
view of reproducing atmospheres of real CP stars. Theoreticians agree that
atomic diffusion must not be considered as acting alone. Although there have
been significant advances in the understanding of the main trends in abundance
peculiarities observed in CP stars -- as for instance the dependence of Mn
overabundances on $T_{\rm{eff}}$ in HgMn stars \citep{AlecianAlMi1981,
SmithSmDw1993r}, or the stratified abundances of Mg, Si, Ca, Ti, Cr, Fe, and Ni
at $8000$\,K and $12000$\,K \citep{LeBlancLeMoHuetal2009l} -- numerical
modelling of atomic diffusion in atmospheres continues to improve only slowly.
Step by step the calculations becomes more sophisticated with the final goal to
provide theoretical metal abundance patterns which can be confronted with
observations of individual stars.

Our most recent progress in numerical modelling of CP star atmospheres has been the
first model of theoretical 3D distributions of Cr and Fe in atmospheres with a
non-axisymmetric magnetic field \citep{AlecianAlSt2017}. In this paper, it was shown
that, even if equilibrium solutions\footnote{Equilibrium solution assumes that local element
  concentrations are such that diffusion velocities are equal to zero everywhere in the
  atmosphere (or equivalently, radiative acceleration  modulus is more or less equal to
  gravity).}
favour higher concentrations of metals in regions where the magnetic field is horizontal
(which, according to some authors, is not observed, see \citealt{KochukhovKoRy2018}),
this is expected to happen in most cases so high up in the atmosphere that it could
hardly be detected. It appears
that the photosphere might rather be dominated by abundance patches due to atomic
diffusion in magnetic fields of non-axisymmetric geometry. At this point one should
however keep in mind that virtually all previous attempts to quantitatively predict
abundance stratifications due to atomic diffusion \citet{AlecianAlSt2017} have used
equilibrium solutions (see for instance \citealt{LeBlancLeMoHuetal2009l}), a method
that might legitimately be questioned. In the past, we have often warned in our papers
against mis- or over-interpretations that can be drawn from equilibrium solutions.
Indeed, equilibrium solutions must be interpreted as mapping only the maximum abundances
that can possibly be supported by the radiation field. In some cases, this may
correspond to the real element distribution, but not so in the majority of cases.
Still, in a number of publications observations have been confronted with equilibrium
results and it has hastily been concluded that theoretical predictions are not
confirmed by observations.

We are convinced that stationary solutions\footnote{Stationary solutions are obtained
by solving the time-dependent continuity equation, and correspond to a constant
non-zero particle flux in time and depth throughout the photosphere. The so called
equilibrium solutions constitute a particular class among the family of stationary
solutions, with particle flux equal to zero.}
obtained with extremely time-consuming calculations of time-dependent atomic diffusion
(which can differ widely from equilibrium results)
will generally be much closer to what happens in real atmospheres; this approach is
discussed in \citealt{AlecianAlStDo2011} and \citealt{StiftStAl2016}. Let us recall
what \citet{AlecianAlSt2017} have emphasised in their conclusions, viz. that the next
step in numerical modelling must consist in the inclusion of stellar mass-loss which
is well known to compete with atomic diffusion in many types of CP stars
\citep[see][]{VauclairVa1975u,MichaudMiTaChetal1983q,BabelBa1992r,AlecianAl1996,
  LandstreetLaDoVa1998,VickViMiRietal2010,AlecianAl2015}. The mass-loss rates, which
have not yet been directly observed for main-sequence CP stars with
  $T_{\rm{eff}} \lta 18000$\,K, have been estimated -- by means of numerical modelling
of stellar internal structure --
by \citet{VickViMiRietal2010} to lie around $5.0\,10^{-14}$ solar mass per year
for hot AmFm stars ($T_{\rm{eff}} < 10000$\,K). X-ray emission
possibly related to the existence of a wind has been detected in an ApBp star
(IQ\,Aur) as discussed by \citet{BabelBaMo1997}. The mass-loss rate probably
increases with effective temperature. For instance, for He-rich stars, which are
the hottest main-sequence CP stars ($T_{\rm{eff}} \gta 18000$\,K), it has been
estimated at about $10^{-12}$ solar mass per year \citep[see][]{VauclairVa1975u}.
Winds are detected through the study of X-ray emission in early
B and hotter stars \citep[see for example][]{OskinovaOsToIgetal2011}.

The present work addresses for the first time the build-up of abundance
stratifications in stellar atmospheres due to time-dependent atomic diffusion in
conjunction with mass-loss. Our study concerns only ApBp stars
(including HgMn stars). In some other CP stars (viz. the AmFm), the mixing of
external layers prevents atomic diffusion of metals to be efficient in the
atmospheres. For stars with $T_{\rm{eff}} \gta 18000$\,K, abundances cannot stratify
significantly in the atmospheres due to the high wind velocity.
Actually, \citet{BabelBa1992r} has been the first to consider a stellar wind
in presence of atomic diffusion in magnetic atmospheres in order to explain the
observed Ca, Cr, Fe and Sr lines in
53 Cam (a cool Ap star with $T_{\rm{eff}}\approx 8500$\,K). This pioneering work
considered equilibrium solutions, assuming a weak inhomogeneous wind as a free
parameter, and proposed a wind model corresponding to a mass-loss rate of about
$3\,10^{-15}$ solar mass per year at the poles to fit the line spectra. In the
present work, we base our findings on time-dependent numerical simulations  for
plane-parallel atmospheres, assuming various mass-loss rates with essentially no
magnetic fields (which may be relevant for HgMn stars) but we also present a
couple of calculations with very weak and with moderate magnetic fields. The
magnetic case accompanied by an anisotropic wind will be considered in a
forthcoming paper.

In Sec.~\ref{modmassloss} we detail how mass-loss is introduced into the equations,
and in Sec.~\ref{numerics} we present the modifications to our code CaratMotion. The
numerical results for a typical Bp star are presented in Sec.~\ref{stratif}.
In Sec.~\ref{caseHD}, we discuss the case of the HgMn star HD\,53929 for which
\citet{NdiayeNdLeKh2018} have derived empirical vertical stratifications
of P and Fe. We finally propose a general discussion in Sec.~\ref{discuss}, followed
by concluding remarks in Sec.~\ref{conc}.

\section{Modelling the mass-loss}
\label{modmassloss}

At present it does not appear feasible to develop a fully self-consistent treatment of
atomic diffusion in combination with mass-loss. However, to a first approximation, the
velocity of the stellar wind (in cgs units) is given by

\begin{equation}
  V_{\rm wind} = \frac{\dot{M}\,M_{\sun}} {4 \pi \rho R^2} =
              \frac{\dot{M}\,g} {4 \pi \rho\,G (M / M_{\sun})}\, ,
\label{eq:wind}
\end{equation}

\noindent
where $\dot{M}$ stands for the mass-loss rate  (solar mass per second). $G$ is the
gravitational constant, $\rho$ denotes the mass density at a given layer in the
atmosphere, $M$ the stellar mass, $R$ the stellar radius, $g$ the stellar gravitational
acceleration. Keeping in mind that the depth of a stellar atmosphere is negligible in
relation to the stellar radius, our formula expresses the fact that the mass flux $V_w
\rho = (dM / dt) / A$ -- with $A$ the stellar surface area -- through the atmospheric
layers remains constant. This approximation has been used several times in the framework
of atomic diffusion modelling \citep[see for
  instance][]{VauclairVa1975u,MichaudMiTaChetal1983q,AlecianAl1986a,VickViMiRietal2010}.
At this stage, we do not need to adopt any hypothesis as to the structure of the wind
(isotropy, homogeneity or time dependence). The mass-loss rate is just a free parameter
fixing the wind velocity according to the depth in our plane-parallel models.

For a given atmospheric model ($T_{\rm eff}$, $\log g$) and corresponding mass $M$, the
formula for the wind velocity becomes
\begin{equation}
V_{\rm wind} = C / \rho. 
\label{eq:Cwind}
\end{equation}
In order to determine $M$ and thence the factor $C$ we resorted to
\citet{SchallerScScMeetal1992} who provide a grid of evolutionary tracks for various
stellar masses. Effective temperature $T_{\rm eff}$ and gravitational acceleration $\log
g$ -- which define our atmospheric models -- yield the stellar luminosity via
\begin{equation}
\log (L / L_{\sun}) = -10.607 - \log g + \log (M / M_{\sun}) + 4\,\log T_{\rm eff} \,.
\label{eq:lum}
\end{equation}
The $\log (L / L_{\sun})$ values for the given stellar $T_{\rm eff}$ and various masses
represented in the evolutionary grid are determined by interpolation, resulting in a
relation $\log g$ vs. $M$. The mass corresponding to the atmospheric parameters is
calculated by interpolation in this relation. For instance, in Sec.\,\ref{stratif},
we use a model atmosphere with $T_{\rm eff} = 12000$\,K and $\log g = 4.0\,$.
Equ.\,\ref{eq:lum} leads to a stellar mass of 3.4 solar masses, hence the factor $C$
in Eq.\,\ref{eq:Cwind} becomes $\approx 1.1\,10^{-14}$ for a mass loss rate of
$10^{-16}$ solar mass per year.

For the star HD\,53929 discussed in Sec.\,\ref{caseHD}, with $T_{\rm eff} = 12700$\,K
and $\log g = 3.71$ \citep{NdiayeNdLeKh2018} we have $C \approx 4.64\,10^{-15}$, again
assuming a mass loss rate of $10^{-16}$ solar mass per year.

\subsection{Numerics}
\label{numerics}

Let us recall that our {\sc CaratMotion} code derives from
{\sc Carat} and {\sc CaratStrat} which calculate radiative accelerations and
diffusion velocities; the physics included in these codes are described in
detail in \citet{AlecianAlSt2004} and \citet{StiftStAl2012}. The numerics of
{\sc CaratMotion} have been presented in \citet{AlecianAlStDo2011}. To ensure
self-consistency between chemical stratifications and atmospheric structure,
the atmosphere is recalculated after each time step, based on the new
abundances, and the atomic line opacities updated accordingly. It comes as
no surprise that time-dependent atomic diffusion can prove very expensive,
a complete run taking up to 300 and more days mono-processor time on a modern
server. Since not all parts of the code can be completely parallelised, it
makes no sense to run it on machines doted with hundreds of processors. Instead,
executing several jobs concurrently on a dedicated 64 core server minimises idle
cpus. Obviously this severely constrains the scope of our simulations which
cannot encompass the bulk of ApBp stars, but is limited to a moderate number of
well chosen cases.

It proved quite straightforward to adapt the numerical scheme for the modelling of
time-dependent atomic diffusion as employed by \citet{StiftStAlDo2013} in order to
accommodate mass-loss. We recall that our models are self-consistent for abundance
stratifications and computed at each time-step with  Kurucz's {\sc Atlas12} code
(\citealt{KuruczKu2005l}, \citealt{Bischof2005}). All chemical elements are transported
with the same positive velocity $V_{\rm wind}$; a positive or negative diffusion velocity
$V_{\rm elem}$ has to be added for the ``diffusing'' elements. As an example, in the case
of HD\,53929 where we are interested in the diffusion of phosphorus and iron only,
the donor-cell scheme has to be applied to 3 elements, with H in addition to P and Fe 
(it will be Mg and Fe in the case discussed in Sec.\,\ref{stratif}). Since abundances
in our numerical code {\sc CaratMotion} are defined with respect to hydrogen one just
has to update the hydrogen number density to obtain at the same time the new number
densities of all the other elements, except P and Fe. The latter are calculated by
substituting the diffusion velocity in the continuity equation by $V_{\rm wind} + V_{\rm P}$
and $V_{\rm wind} + V_{\rm Fe}$ respectively.

\section{Build-up of abundance stratifications assuming mass-loss}
\label{stratif}

The abundance stratifications build-up assuming a mass-loss compared to the case without
mass-loss \citep[as recently discussed by][]{StiftStAl2016}, essentially differs by the
fact that the stationary solution can hardly converge to equilibrium (zero, or very small
particle flux), except for elements that are very weakly supported in the upper atmosphere
by the radiation field \citep[such as for helium, see][]{VauclairVa1975u}. That is due to
the fact that mass-loss imposes a positive flux of particles even if diffusion velocity is
close to zero. Therefore, the presence of mass-loss facilitates the escape of overabundant
metals from the atmosphere and tends in most cases to produce lower vertical abundance
stratification contrasts.
In some cases, as for He previously mentioned, mass-loss can compensate
gravitational settling (negative diffusion velocity) and make elements overabundant that
would sink towards deeper layers in the absence of mass-loss. Detailed stratifications
depend on the relative strength and sign of diffusion and wind velocities at each depth
point, When the mass-loss rate is too high (generally it increases with effective
temperature), atomic diffusion is no longer able to produce abundance stratifications,
since matter flux from deeper layers erases abundance changes faster than atomic diffusion
can produce them. In this case, the star presents \emph{normal} abundances. This is
believed to be the cause of the higher $T_{\rm eff}$ cutoff of the CP phenomenon.

Our numerical code {\sc CaratMotion} models the time-dependent abundance stratification
process in a realistic, strongly non-linear way, from a starting time when abundances
are generally assumed to be vertically homogeneous and solar -- with non-constant
particle flux throughout the atmosphere -- to a final time when particle fluxes become
constant and abundances stratified. Thereafter, stratifications no longer evolve
over the timescales considered (a small fraction of the main-sequence life, see below);
this is what we call a stationary solution. Even though the numerical process is expensive
and necessitates large amounts of cpu time, quite often the time-dependent diffusion
calculations converge to a stationary solution after a reasonable number of time steps,
corresponding to a very short physical time (from some tens of years to some thousands,
depending on the element) compared to the characteristic evolutionary time for the
internal structure (or stellar age). Hereafter, we present only the final stationary
solutions, assuming that they correspond to the observed abundance stratifications,
transient phases having a very small chance of being observed.

\subsection{Abundances of Mg and Fe in non-magnetic atmospheres}
\label{noB}

\begin{figure*}
\centering
\includegraphics[width=170mm]{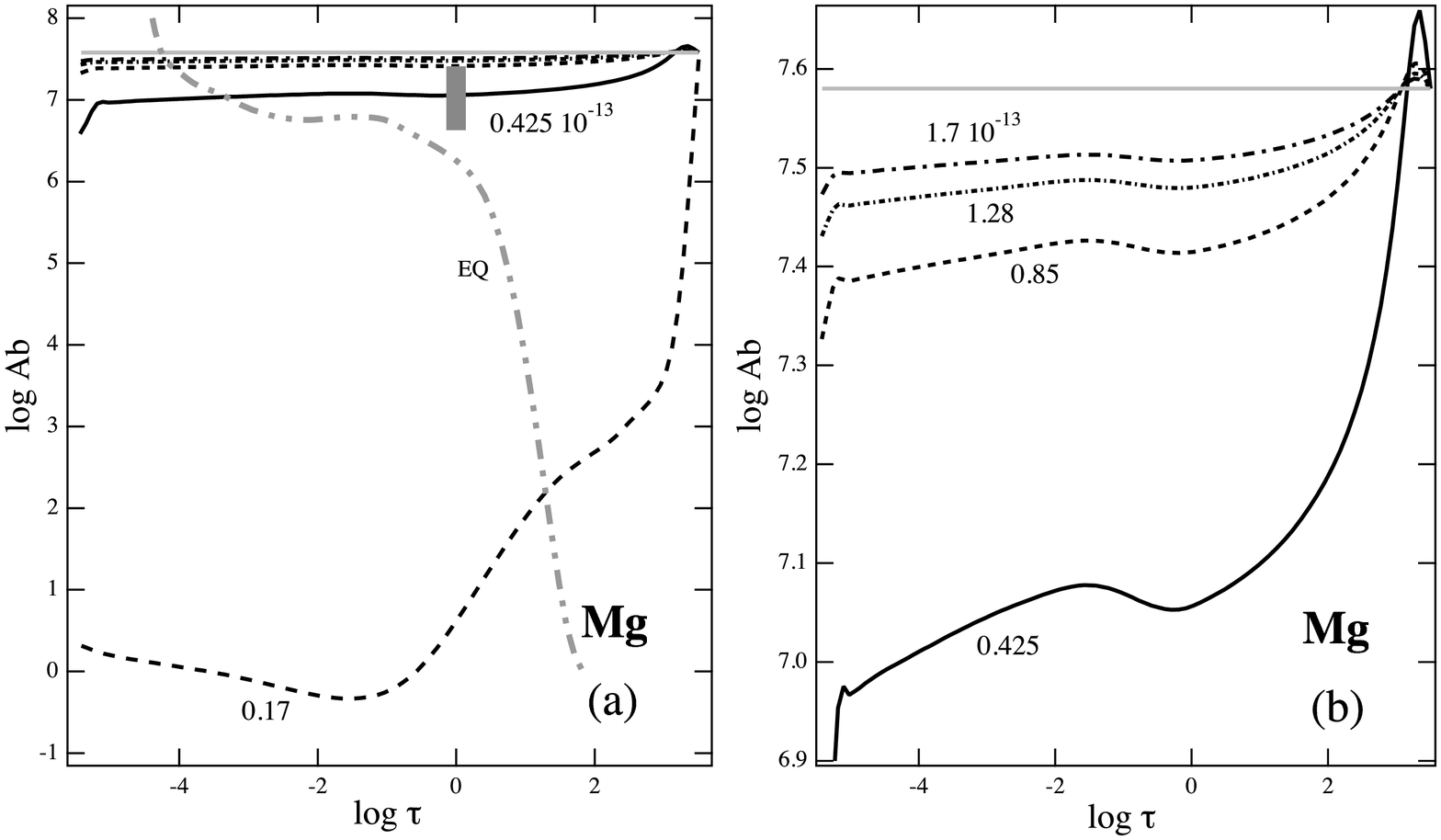}
 \caption{
   Mg abundance stratification (stationary solutions) vs. the logarithm of the optical
   depth at 5000\,{\AA} as a function of various mass-loss rates
   (model with $T_{\rm{eff}}= 12000$\,K, and $\log{g}= 4.0$).  Panel (a) shows the Mg
   stratifications for 5 different mass-loss rates (1.7, 1.28, 0.85, 0.425, 0.17 in
   units of $10^{-13}$ solar mass per year). The vertical thick grey bar corresponds
   to observations of Mg overabundances in HgMn stars having approximately the same
   $T_{\rm{eff}}$ and $\log{g}$~. The equilibrium solution (EQ) -- heavy dash-dot-dot
   grey line -- is also shown (see text). Panel (b) presents the same stratifications
   as panel (a) except for the lowest mass-loss rate, but with a different vertical
   scale.
 }
\label{fig:Layout_Mg}
\end{figure*}

HgMn stars are considered non-magnetic CP stars and should therefore be easier to
model. Some authors \citep{MathysMaHu1995} have claimed detection of a magnetic
field in one HgMn star, but this has not yet been confirmed. On the other hand,
\citet{AlecianAl2013l} proposed that a weak magnetic field could explain the possible
existence of high altitude spot-like clouds in HgMn stars. In this section, we assume
that HgMn stars are strictly non-magnetic.

Looking at the abundance anomalies in HgMn stars, it appears that the abundance
of Mg is very often close to the solar value or slightly underabundant
\citep[see the compilation of][]{GhazaryanGhAl2016}. However, since according
to \citet{AlecianAl2015} Mg is only weakly supported by the radiation field in
the line forming region, it should be systematically strongly underabundant 
when diffusion is supposed to act alone. Therefore it is interesting to test
the role of mass-loss for this element.

\begin{figure}
\includegraphics[width=85mm]{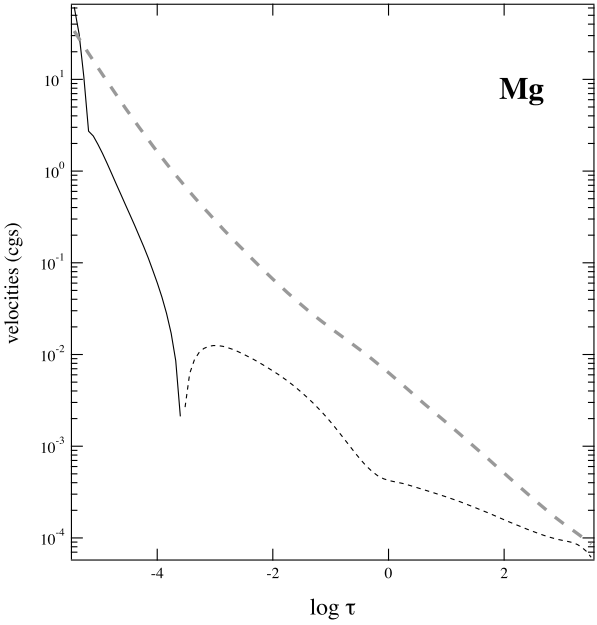}
 \caption{Mg diffusion and wind velocity respectively for the stationary solution. The solid
          line stands for positive diffusion velocities (upwards), the short-dashed line for
          negative velocities. The long-dashed line shows the wind velocity for a mass-loss
          rate of $0.425\,10^{-13}$ solar mass per year.}
\label{fig:Layout_Mg_vit}
\end{figure}

\begin{figure}
\includegraphics[width=85mm]{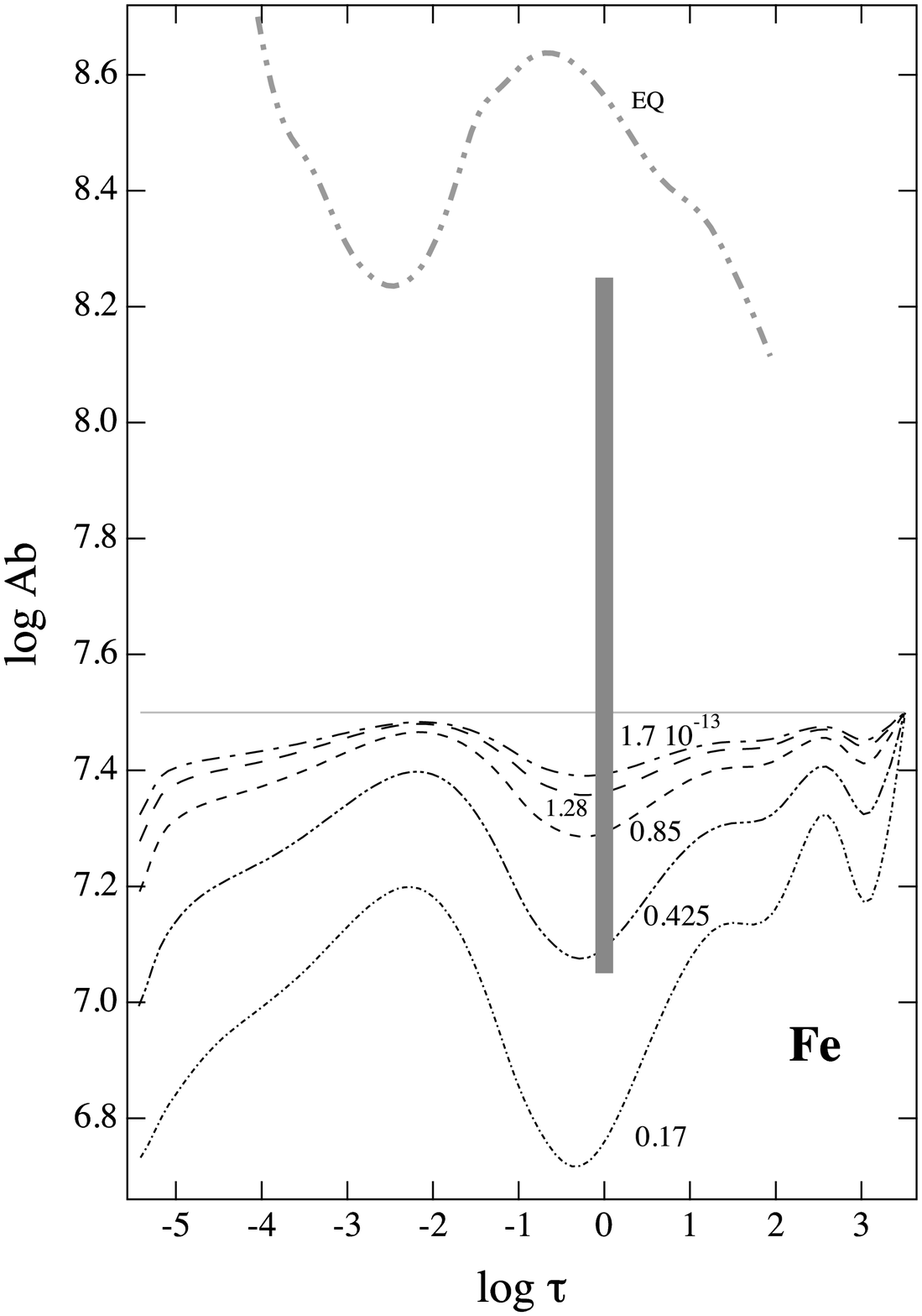}
 \caption{Same as Fig.\,\ref{fig:Layout_Mg}a, for Fe.}
\label{fig:Layout_Fe}
\end{figure}

\begin{figure}
\includegraphics[width=85mm]{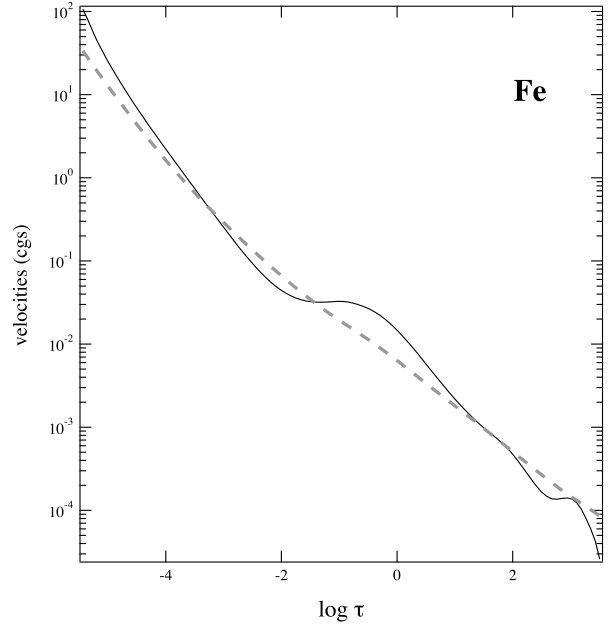}
 \caption{Same as Fig.\,\ref{fig:Layout_Mg_vit} for Fe.}
\label{fig:Layout_Fe_vit}
\end{figure}

Stationary solutions for various mass-loss rates in an atmosphere with
$T_{\rm{eff}}= 12000$\,K, and $\log{g}= 4.0$ are shown in Figs.\,\ref{fig:Layout_Mg}
to \ref{fig:Layout_Fe_vit}. Since a global mass flow (here the wind) reduces
vertical abundance contrasts, it improves the numerical stability of the
calculations. Thanks to the wind, many calculations which become numerically
unstable (or do not converge within some reasonable cpu time) in the absence of
mass-loss now become possible\footnote{Numerical stability and convergence of
{\em all} diffusing elements have to be met, at least over those layers that
mainly contribute to the spectral line formation, before we consider a stationary
solution of time-dependent diffusion well enough established.}. 
In this study, we have succeeded in calculating stationary solutions for Mg and
Fe when mass-loss rates are higher than $0.17\,10^{-13}$ solar mass per year.
Although this lower limit seems rather far removed from the zero mass-loss case
which we would also have liked to model, our results appear to justify the claim
that lower mass-loss rates are unlikely to fit the majority of observed abundances
of Mg and Fe (see our discussion in Sec.\,\ref{thegauge}).

On account of the time-consuming nature of our numerical
calculations, we cannot afford to explore the domain of possible mass-loss
rates and their effect on the various chemical elements in much detail. In
a first step, we have chosen 5 values starting from the lowest rate of
$0.17\,10^{-13}$ solar mass per year mentioned above.
In Fig.\,\ref{fig:Layout_Mg}a, stationary solutions for Mg are shown for all 5
mass-loss rates considered. As previously explained, Mg is not well supported by
the radiation field, becoming therefore extremely depleted in the atmosphere for
the lowest mass-loss rate of
$0.17\,10^{-13}$ solar mass per year. Attention should be paid to the fact that
the equilibrium solution (dash-dot-dot grey curve labelled EQ) is very different
from any of the stationary solutions. Fig.\,\ref{fig:Layout_Mg_vit} displays the
diffusion velocity together with the wind velocity for a mass-loss rate of
$0.425\,10^{-13}$. Looking at the stationary solution, the diffusion velocity turns
out to be significantly lower than the wind velocity almost throughout the atmosphere.

Fig.\,\ref{fig:Layout_Fe} and \ref{fig:Layout_Fe_vit}, present the stationary solutions
for Fe (same mass-loss rates as for the Mg results). This time, the diffusion velocity
is always positive and in closer competition with the wind velocity.

\subsection{Abundances of Mg and Fe in magnetic atmospheres}
\label{withB}

\begin{figure}
\includegraphics[width=85mm]{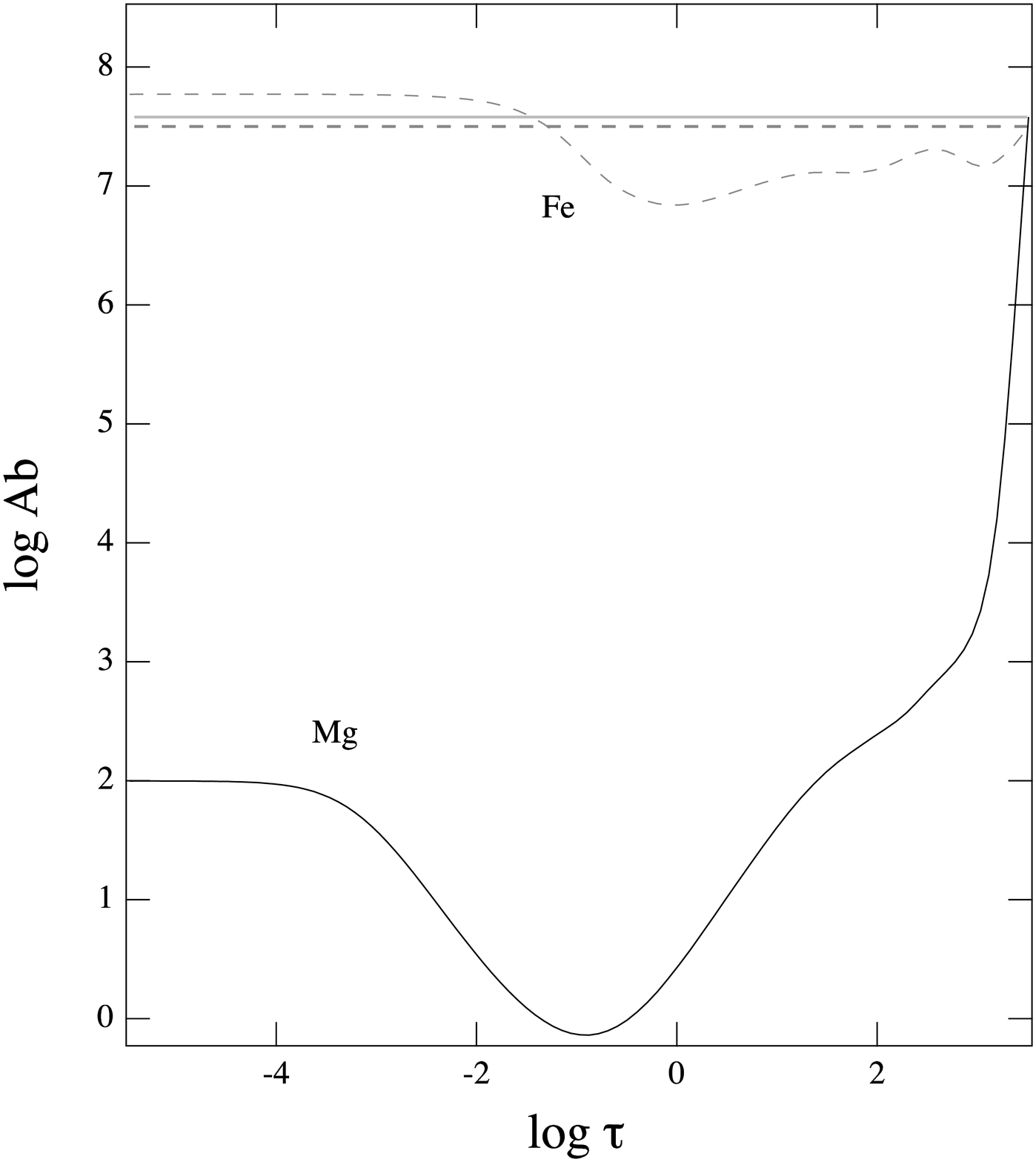}
\caption{Stationary solutions in the same atmospheric model as for
  Fig.\,\ref{fig:Layout_Mg}, but with a horizontal magnetic field of 5\,kG
  (solid  lines for Mg, dashed ones for Fe). The mass-loss rate is $0.17\,10^{-13}$,
  the smallest one for which the solution converges. The dashed light-grey line gives
  the Fe stationary solution, the solar abundance is plotted in dark-grey.}
\label{fig:FIG_Strat_finales_magnB}
\end{figure}

\begin{figure}
\includegraphics[width=85mm]{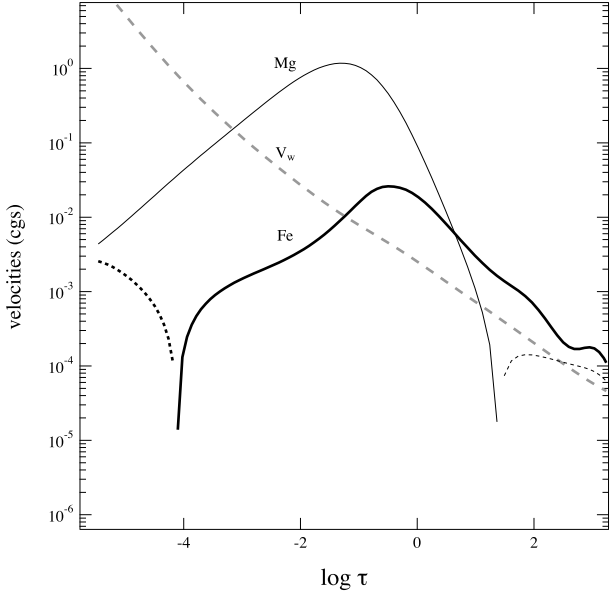}
 \caption{Diffusion velocities of Mg (thin black lines) and Fe (heavy black lines) for the
  stationary solutions of Fig.\,\ref{fig:FIG_Strat_finales_magnB}, and wind ($\rm{V_w}$)
  velocity (grey dashed line).}
\label{fig:FIG_vit_finales_B_5kG90d1m14}
\end{figure}

We have carried out a number of calculations for a magnetic atmosphere. However,
these results -- obtained for a plane-parallel model -- have to be considered with
extreme caution since one may reasonably suppose that in the magnetic case the
stellar wind should be anisotropic \citep{BabelBa1992r}. It is well known that the
diffusion velocity does not directly depend on the vertical component of the magnetic
field vector\footnote{The (indirect) sensitivity of the diffusion velocity to the
vertical component of the magnetic field arises from the Zeeman effect which increases
the radiative acceleration \citep[see][]{AlecianAlSt2004}.}, but that it depends
instead explicitly on the horizontal component of the field. To illustrate the effect
of a magnetic field, we have considered the case of a horizontal field of moderate
strength (5\,kG). We expect that the wind will be impeded by horizontal magnetic lines,
similarly to what happens to the diffusion velocity (but certainly in a different
way\footnote{The diffusion velocity is closely related to the collision time between ions
and protons. It is a microscopic process, whereas the wind is a macroscopic-hydrodynamic
process.}) since atoms are mostly ionised in our models. Establishing stationary solutions
for Mg and Fe using the same model as in the previous section, we therefore adopted the
lowest mass-loss rate ($0.17\,10^{-13}$) for which the solution converges. The results are
shown in Fig.\,\ref{fig:FIG_Strat_finales_magnB} and \ref{fig:FIG_vit_finales_B_5kG90d1m14}.
It transpires that in the uppermost layers, stationary stratifications are very different
from the non-magnetic ones; diffusion velocities display a completely different profile
over a large part of the atmosphere.

\begin{figure}
\includegraphics[width=85mm]{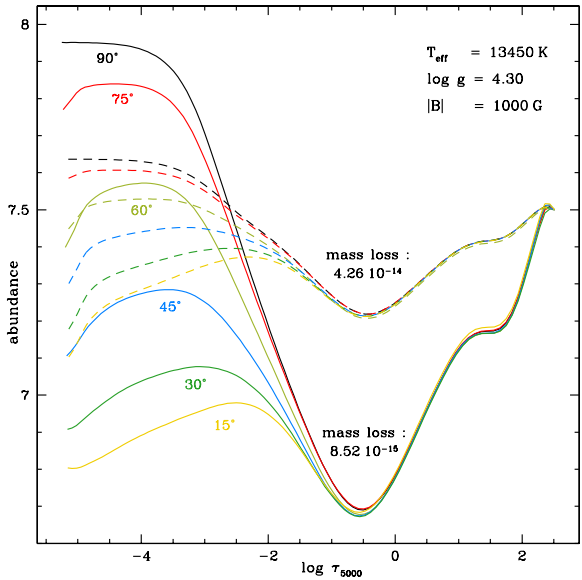}
\caption{Fe abundance profiles as a function of optical depth, angle, and mass loss
  in an atmosphere with $T_{\rm eff} = 13450$\,K and $\log g = 4.30$. A magnetic field
  strength of $|B|= 1000$\,G is assumed at angles $15^{\circ}, 30^{\circ}, 45^{\circ},
  60^{\circ}, 75^{\circ}, 90^{\circ}$ relative to the surface normal. Full lines refer
  to a mass-loss rate of $8.5\,10^{-15}$ solar mass per year, dashed lines to
  $4.26\,10^{-14}$ solar mass per year.}
\label{fig:t13450_angles}
\end{figure}

It is not without interest to have a look at the relative impact of
magnetic fields and of mass loss on final stationary abundance distributions.
Taking a stellar atmosphere with $T_{\rm eff} = 13450\,$K and $\log g = 4.30$ --
similar to the one derived by \citet{Castelli_et_al_2017} for HR\,6000 -- a field
strength of 1000\,G and mass loss rates of $8.5\,10^{-15}$ and $4.26\,10^{-14}$
respectively, we arrive at the angle-dependent stratifications shown in
Fig.\,\ref{fig:t13450_angles}. As one would expect, the maximum difference in
abundance between the horizontal and the vertical field case decreases with
increasing mass loss. Whereas at $8.5\,10^{-15}$ abundances at $90^\circ$ are higher
by about 1.15\,dex than at $15^\circ$, this difference drops to about 0.55\,dex when
the mass loss is $4.26\,10^{-14}$. For field angles up to $45^\circ$, the solutions for
$4.26\,10^{-14}$ lie consistently (by up to 0.5\,dex) above those for $8.5\,10^{-15}$
throughout the atmosphere; when the angles exceed $60^\circ$, this is only true in
those parts of the atmosphere where the magnetic field plays no role (below about
$\log \tau = -0.5$).

At this point it is important to discuss an aspect of diffusion in
magnetic stellar atmospheres that has received only the scantest attention in the
past but whose repercussions on the comparison diffusion theory vs. empirical
non-homogeneous abundances as derived from Zeeman Doppler mapping are of utmost
relevance. We want to stress that all the exploratory modelling of the build-up of
abundance stratifications in magnetic CP stars so far has employed a 1D approach,
looking at the temporal evolution of the atmospheric structure in a ``cylinder'' of
material isolated from its surroundings. In other words, only vertical velocities
have been allowed, and the ``cylinder'' is assumed to stay in perfect pressure
equilibrium with its surroundings all the time. On the other hand however, as
Fig.\,\ref{fig:t13450_angles} reveals, different field angles lead to different
stratifications and to different atmospheric structure. From Fig.\,9 of 
\citet{Stift_Leone_2017a} one can deduce that angle-dependent stratifications
will lead to angle-dependent horizontal pressure differences between adjacent
``cylinders''. In the absence of an adequate stabilising force -- e.g. a strong
{\bf vertical} magnetic field -- horizontal pressure equilibrium will be
established almost instantaneously, resulting in mixing between the
``cylinders''. The problem has suddenly become 3D, even in the simple case of
a centred dipole geometry. Although in a first approximation such mixing occurs
only horizontally, independently for each geometric height in the atmosphere,
it is clear that the resulting density, pressure and temperature structures of
the individual ``cylinders'' will not in general ensure hydrostatic equilibrium.
Relaxing to hydrostatic equilibrium in each cylinder should again entail deviations
from horizontal pressure equilibrium. Whereas a magnetic star should adapt to
the abundance-buildup in the shortest of times, the problem gets highly complex
for any kind of numerical approach. Any realistic 3D modelling of atomic diffusion
in strongly magnetic CP stars will have to deal with this additional challenge.

How does Zeeman Doppler mapping enter this discussion? Virtually
every single abundance map in the literature has been derived under the assumption
that the horizontal distribution of the chemical elements may be non-homogeneous
but that all abundances, wherever on the stellar surface, have to be unstratified,
i.e. constant from top to bottom of the atmosphere. In addition, in order to
constrain the solutions to this ill-posed 2D inversion problem, people have
mostly resorted to regularisation functions which look for the smoothest
possible map. This approach is at variance with basic stellar astrophysics:
Fig.\,9 of \citet{Stift_Leone_2017a} makes it abundantly clear that unstratified
but horizontally non-homogeneous abundances structures cannot be in horizontal
pressure equilibrium, independently of the approximations used in time-dependent
atomic diffusion modelling. An approach by \citet{Ruso2016}
for 3D abundance mapping with the help of step-like abundance profiles that vary
with position on the star also fails to reflect these and other (astro-)physical
facts. The local profile in Rusomarov's thesis is defined by the abundance in the
high atmospheric layers $\epsilon_{up}$, the abundance deep in the atmosphere
$\epsilon_{lo}$, the position $d$ of the transition region and its width $\delta$
where the abundance changes between $\epsilon_{up}$ and $\epsilon_{lo}$ (see his
Fig.\,2.3). Numerical models have established a bewildering multitude of vertical
abundance profiles that frequently bear little resemblance to a step function as
e.g. shown in Fig.\,\ref{fig:t13450_angles}. The position $d$ of the transition
region always changes with magnetic field strength, the optical depth where the
abundance reaches its maximum depends on the field angle. the spread in
$|\epsilon_{up} - \epsilon_{lo}|$ is a function of field strength and field angle.
Rusomarov's regularisation function minimises all conceivable differences, i.e.
between the horizontal gradients of the widths $\delta$, of the positions $d$, of
the upper abundances $\epsilon_{up}$, and of the lower abundances $\epsilon_{lo}$.
Additionally, the spread in $|\epsilon_{up} - \epsilon_{lo}|$ has to be minimum (see
his formula 2.16), but all this is at variance with what we know about atomic
diffusion in magnetic stellar atmospheres. There is no known/feasible physical
mechanism that would lead to the kind of stratification plotted in Figs.\,4 and 8
of \citet{RusoKochRyab2016}.

To put it succinctly, neither are there empirical abundance
stratifications of strongly magnetic CP stars available that could be confronted
with theoretical results, nor is theory presently capable of treating the full
3D nature of diffusion in these stars. When it comes to real stars, in the present
study we therefore prefer to stick to those which are non-magnetic or only weakly
magnetic.

\section{The case of HD\,53929}
\label{caseHD}

\begin{figure}
\includegraphics[width=85mm]{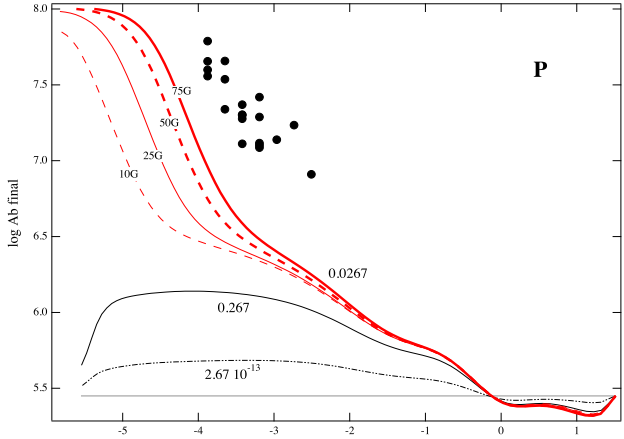}
 \caption{
Stratification of P in the atmosphere of HD\,53929. The filled circles show the abundances
empirically determined by \citet{NdiayeNdLeKh2018} from observations. The solid grey
line is the solar abundance, the thin solid and dash-dot-dot curves are our stationary
non-magnetic solutions for respective mass-loss rates of 2.67 and 0.267$\,10^{-13}$ solar
mass per year. The group of red curves labelled 10\,G, 25\,G, 50\,G, and 75\,G are the
solutions obtained by assuming a weak horizontal magnetic field, and for the lowest
mass-loss that ensures convergence of this model (0.0267$\,10^{-13}$).
 }
\label{fig:FigMNRAS_P_12750}
\end{figure}

\begin{figure}
\includegraphics[width=85mm]{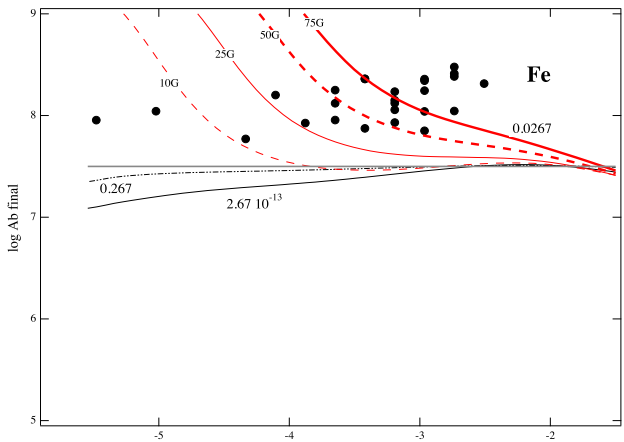}
 \caption{
 Same as Fig.\,\ref{fig:FigMNRAS_P_12750} for Fe.
 }
\label{fig:FigMNRAS_Fe_12750}
\end{figure}

Recently, \citet{NdiayeNdLeKh2018} have published an abundance analysis of 2 HgMn stars
(HD\,53929 and HD\,63975). These authors found well marked abundance stratifications of
phosphorus, and slight iron overabundances with very moderate abundance gradients. Both
stars exhibit similar trends in the respective P and Fe abundance stratifications. We
decided to apply our time-dependent atomic diffusion modelling to one of these stars.
We have chosen HD\,53929 which seems a less evolved main sequence star than HD\,63975,
thus a more typical HgMn star. According to \citet{NdiayeNdLeKh2018}, the atmospheric
parameters are $T_{\rm{eff}}\approx 12750$\,K and $\log{g}\approx 3.7$. The empirical
abundance stratifications of P and Fe are shown in Fig.\,\ref{fig:FigMNRAS_P_12750}
and \ref{fig:FigMNRAS_Fe_12750} respectively (filled black circles); we have used the
same axis scales as in their paper.

The curves plotted in Fig.\,\ref{fig:FigMNRAS_P_12750} and \ref{fig:FigMNRAS_Fe_12750}
correspond to the stationary solutions of our numerical simulations for various magnetic
and mass-loss parameters. In these calculations, P and Fe are allowed to diffuse
simultaneously, the atmospheric models being kept self-consistent by recomputing them
with the actual abundances of both P and Fe at each time step -- as we did previously
for Mg and Fe (Sec.\,\ref{numerics}). The other elements are assumed to retain their
solar ratios relative to hydrogen.

We first considered the non-magnetic case with various values of the mass-loss
rate. The lowest mass-loss rate ($0.267\,10^{-13}$) differs from the lowest one
of the model used in Sec.\,\ref{stratif} (for non magnetic case) because of the
higher effective temperature and lower gravity of HD\,53929. The highest
mass-loss we show here is 10 time larger. For both elements, the stationary
solutions are far from the stratifications determined by
\citet{NdiayeNdLeKh2018}. In particular, in high layers P remains marginally
overabundant, Fe depleted. This may be explained by the fact that radiative
accelerations largely exceed gravity in these layers, helping P and Fe to escape
the star. To impede this escape one must assume a horizontal magnetic field.
This magnetic field however has to be weak enough (or not well structured on a
large scale) to be compatible with the lack of detection in HgMn stars (see the
discussion in Sec.\,\ref{discuss}). In our calculations we have adopted weak
horizontal fields in the range from 10\,G to 75\,G and the lowest mass-loss rate
that yields stable stationary solutions (see Sec.\,\ref{withB}). This mass-loss
rate of $0.0267\,10^{-13}$ solar mass per year is small enough in our opinion to
be assimilated to the zero mass-loss case\footnote{We make the assumption that
horizontal magnetic fields block the wind.}. Results are shown in
Fig.\,\ref{fig:FigMNRAS_P_12750} and \ref{fig:FigMNRAS_Fe_12750}, the curves
labelled with the magnetic field strength. Since the diffusion velocity drops
drastically in higher layers due to the horizontal field, elements cannot easily
escape the atmosphere but accumulate in these layers. The stronger the field,
the closer the phosphorus stratification gets to those empirically determined
from observations. The situation is less convincing for the iron stratification
profile; the calculated overabundances however are now closer to the empirical
values.

\section{Discussion}
\label{discuss}

\subsection{Abundance of Mg in HgMn stars: a mass-loss rate gauge}
\label{thegauge}

As stressed in Sec.\,\ref{noB}, Mg is an interesting element for constraining models,
since it is observed moderately underabundant in HgMn stars whereas it should be strongly
underabundant (by several orders of magnitude!) according to numerical simulations with
atomic diffusion acting alone -- no mass-loss -- in non-magnetic atmospheres. From our
results shown in Fig.\,\ref{fig:Layout_Mg} it appears that adding mass-loss to atomic
diffusion may explain observed abundances. These abundances have been taken from the
compilation of \citet{GhazaryanGhAl2016} and are indicated by the vertical thick grey
bar at $\log \tau = 0.0$ in Fig.\,\ref{fig:Layout_Mg}a. It turns out that the stationary
solutions for Mg are very sensitive to the mass-loss rate adopted in the simulations.
For the $T_{\rm{eff}} = 12000$\,K, $\log{g} = 4.0$ model, and among the five different
mass-loss rates we have considered, 0.425$\,10^{-13}$ solar mass per year yields the
stationary solution best in accord with the observed abundances of Mg. The same
mass-loss leads to stationary solutions for Fe compatible with the observed abundances.
Of course we cannot and do not claim that the mass-loss rate that has to be adopted for
this model should be exactly 0.425$\,10^{-13}$ solar mass per year. It seems clear however,
that mass-loss rates of about 0.17$\,10^{-13}$ and lower are incompatible with the
observed abundances.

Since Mg is only weakly supported by radiative acceleration at solar abundance (the
radiative acceleration becomes lower than $-g$ for $\log\tau>{-4}$), the depletion
of Mg is due to gravitational settling throughout the major part of the atmosphere,
including deep layers where the diffusion velocity is no longer sensitive to magnetic
fields. This means that even in the presence of a strong horizontal magnetic field,
depletion of Mg will occur in deep layers; this is confirmed by observations which
show that Mg is often depleted in magnetic ApBp stars. We are therefore inclined to
consider magnesium  a very good mass-loss indicator in hot CP stars.

\subsection{The case of Fe}
\label{iron}

The vertical thick grey bar at $\log \tau = 0.0$\,\, in Fig.\,\ref{fig:Layout_Fe}
marks the range of observed iron abundances in the HgMn stars considered in
Fig.\,\ref{fig:Layout_Mg}a. We can see that the abundances of Fe range go from
-0.5 \,dex below to about +0.8\,dex above the solar value. According to our results,
the lowest mass-loss rate of 0.17$\,10^{-13}$ may be excluded, but all the others
(including 0.425$\,10^{-13}$) are found compatible with observations. Whatever the
overabundance, to obtain it one has to assume the presence of horizontal magnetic
fields, even
though they might be weak ones, as in the case of HD\,53929 (see also
Fig.\,\ref{fig:FIG_Strat_finales_magnB}).

\subsection{Confronting time-dependent models with abundance stratifications
  deduced from observations}
\label{confronting}

We have considered in detail the case of HD\,53929, establishing stationary
solutions for P and Fe. During the calculation of the build-up of the abundance
stratifications, the atmospheric structure had to adapt continuously; effective
temperature and gravity taken from \citet{NdiayeNdLeKh2018} were kept constant.
Our results for phosphorus are rather satisfactory since the abundance gradient
of P is very close to the one determined empirically by these authors. Our best
solution (with a horizontal magnetic field of 75\,G) lies by about 0.5\,dex
below the empirical stratification, which is not really troubling, given the
overabundance of 2.5\,dex. Notice that the results shown in
Fig.\,\ref{fig:FigMNRAS_P_12750} for P may be compared to the stratification
found by \citet{CatanzaroCaGiLeetal2016}. These authors have found a deeper
stratification of P in a HgMn star (HD\,49606) with effective temperature and
gravity not very different from HD\,53929. In their work P is more enhanced than
in our result, but the abundance gradient is compatible with the one we have
found for P for $\log \tau > -2.0$. At this depth, according to our
calculations, the stratification does not depend directly on a possible magnetic
field.

Results are less satisfactory for Fe for which we find
decreasing abundances with increasing depth, whereas a weak increase in abundance has
been found by \citet{NdiayeNdLeKh2018}. The values of the predicted Fe overabundances
are, however, quite consistent with the observational values. Please note at this
point that \citet{NdiayeNdLeKh2018} do not consider horizontal inhomogeneities in
their abundance analysis but only vertical ones, whereas we implicitly assume that
horizontal inhomogeneities exist, as soon as magnetic fields are present.

We want to point out an important aspect of our results, viz. that the best solutions
require the existence of a weak horizontal magnetic field. Mass-loss does not seem to
be the determining parameter for both P and Fe, which is at variance to what we find
for Mg (Sec.\,\ref{noB} and \ref{thegauge}). The question arises whether the field
strengths in the calculations for HgMn stars can be considered compatible with the
observations? \citet{AlecianAl2012, AlecianAl2013l} has discussed the consequences of
assuming weak magnetic fields in HgMn stars. He suggested that a weak magnetic field
helps in explaining the presence of spots detected in some HgMn stars for elements
like Nd, Pt, Hg and a few others
\citep{AdelmanAdGuKoetal2002l,KochukhovKoPiSaetal2005,HubrigHuGoSaetal2006f}, without
contradicting the ``canonical'' diffusion model in which metals with high solar
abundances should stratify in an isotropic way deeper in the atmosphere. From our
present results we see that a weak magnetic field could also help in understanding
abundance stratification of metals like phosphorus, perhaps also iron. The existence
of such weak fields presently remains a hypothesis which needs confirmation by
observations.

\section{Conclusion}
\label{conc}

We have extended the applicability of our 1D numerical models
of magnetic and non-magnetic CP star atmospheres by adding  a
mass-loss to atomic diffusion processes. In our calculations, the mass-loss
rate is a free parameter that translates into a mass flow (or
wind) with a given constant flux with respect to the depth. Mass-loss rates of
around $5.0\,10^{-14}$ solar mass per year -- already considered for AmFm stars
by \citet{AlecianAl1996} and \citet{VickViMiRietal2010} -- have been explored.
Our results confirm that abundance stratifications due to atomic diffusion are
indeed very sensitive to the mass-loss rate. We have shown that for a model
atmosphere with $T_{\rm{eff}} = 12000$\,K and $\log{g}=4.0$, Mg will be
extremely depleted in the absence of mass-loss, but actually Mg is found only
mildly underabundant in HgMn stars. Such mild underabundances in our numerical
simulations can only be achieved by assuming a mass-loss rate very close to
$4.25\,10^{-14}$ solar mass per year, a rate that is also compatible with the
observed abundances of Fe in the same class of stars. We
propose to take Mg abundances in HgMn star as a kind of gauge for measuring the
mass-loss rates of these stars. For the near future we are planning to explore
models of various effective temperatures and gravities in order to find out how
mass-loss rates might depend on stellar fundamental parameters.

We have also compared our numerical models to empirical abundance stratifications
deduced from observations of real stars. For this purpose we have chosen the
results of \citet{NdiayeNdLeKh2018} obtained for P and Fe in the HgMn star
HD\,53929. The result for the stratification of P is found rather satisfactory,
but only if a weak (75\,G) horizontal magnetic field is accompanied by a very
small mass-loss rate. This conclusion must not be generalised; we do not
claim that fields of similar strength exist in all HgMn stars. 

Many more
numerical simulations have to be compared to empirical stratifications in
order to get a more complete picture of the interplay of mass-loss and weak
magnetic fields but there can be little doubt that mass-loss is an essential
ingredient to diffusion models of CP stars.

\section*{Acknowledgements}

All codes that have been used to compute the models have been compiled
with the GNAT GPL Edition of the Ada compiler provided by AdaCore; this
valuable contribution to scientific computing is greatly appreciated. This
work has been supported by the Observatoire de Paris-Meudon in the framework
of \emph{Actions F\'ed\'eratrices Etoiles} and partly performed using HPC
resources from GENCI-CINES (grants c2016045021, c2017045021). The authors want
to thank Dr. G{\"u}nther Wuchterl, head of the ``Verein Kuffner-Sternwarte'',
for the hospitality offered.




\bibliographystyle{mnras}
\bibliography{mass_loss} 



\bsp	
\label{lastpage}
\end{document}